\documentclass[11pt,a4paper]{article}

\usepackage[hidelinks]{hyperref}
\usepackage{amssymb, amsmath, graphicx, subcaption}
\usepackage[ruled]{algorithm2e}
\usepackage{multirow}
\usepackage{float}
\usepackage{fancyhdr}

\fancypagestyle{firststyle}
{
	\fancyhf{}
	\fancyhead[RE,LO]{}
	\fancyfoot[RE,LO]{}
	\fancyfoot[LE,RO]{Page \thepage}
}

\begin{document}
	
	\title{Linearized model for satellite station-keeping \\ and tandem formations under the \\ effects of atmospheric drag}
	
	\author{David Arnas\thanks{Massachusetts Institute of Technology (MIT), Cambridge, MA, 02139, USA. Email: \textsc{arnas@mit.edu}}}
	
	\date{}	
	
	\maketitle
	
	\thispagestyle{firststyle}
	
	\begin{abstract}
		This work introduces a linearized analytical model for the study of the dynamic of satellites in near circular orbits under the effects of the atmospheric drag. This includes the evaluation of the station keeping required for each satellite subjected to a control box strategy, and also the study of the dynamic of tandem formations between two or more satellites that are located on the same nominal space-track. The model takes into account the effect of the orbit perturbation provoked by the atmospheric drag, while the effects of the Earth gravitation potential are included in the definition of the nominal orbits of the satellites. This allows to easily define the maneuvering strategies for the satellites involved in the tandem formation and study their absolute and relative dynamic. In particular, this work focuses on the study of a master-slave scenario and the in plane maneuvers that these satellites require, proposing two different control strategies for the formation.
	\end{abstract}

\section{Introduction}

The space sector has experienced an important evolution in the last decades providing a wide number of applications including Earth observation, telecommunications, Earth positioning, defense or research. One of the reasons for this is that satellites provide an unparalleled position to perform their observations, allowing to observe vast regions of the Earth surface in a short period of time, a very difficult task to achieve with technical means in ground. This situation has allowed not only the increase on the number of space missions, but also the possibility to launch satellites subjected to a smaller budget. In that respect, one the most successful design philosophies currently in use is satellite formations. 

Satellite formations are groups of satellites that present a coordinated control during their dynamic. This allows for instance to combine the scientific data of different missions, or to cooperate between satellites to achieve a common task. This kind of design has already been successfully used in several missions such as A-Train~\cite{atrain}, TanDEM-X~\cite{tandemx} and Tanem-L~\cite{tandeml}. Another interesting use of satellite formations is to allow some missions to benefit from the measurements of satellites already in orbit. This allows to significantly reduce the costs of the mission while increasing the number of possibilities in design. Examples of this kind of space missions include FLEX~\cite{carbon}, SAOCOM-CS~\cite{saocom} or SESAME~\cite{sesame}.

This manuscript focuses on the study, under the perturbation produced by the atmospheric drag, of tandem formations between two or more satellites that are located in near circular Low Earth Orbits (LEO). In that respect, this work takes into account the fact that satellites may have different physical properties, that is, their masses, cross section areas or drag coefficients could be different. This means that, in general, the ballistic coefficients of the satellites considered could be different, making the orbit decay that each satellite experiences also different. Therefore, a relative dynamic that strongly depends on the ballistic coefficient ratio between both satellites is generated.

In order to perform this study, this manuscript describes a simple analytical methodology that allows to compute the order of magnitude of the orbital maneuvers required to compensate the effects of the atmospheric drag in a given orbit. This methodology allows an easy understanding and calculation of the problem considered, while obtaining a good accuracy in the estimations. This formulation is then applied to the relative motion between satellites in tandem formation, where it is assumed that the satellites are located over the same trajectory in the ECEF (Earth Centered Earth Fixed) frame of reference. In that sense, the manuscript focuses on a master-slave scenario as an example of direct application of the methodology presented.

Compared to other studies about relative motion and linearized dynamics~\cite{vadali,schaub,clohessy,gim,carter}, this work presents a much simpler set of equations for the study of the problem at the cost of some accuracy. Nevertheless, the proposed model allows to obtain a first order approximation of the dynamic of the system that can be used to clearly identify the effects of the variables involved in the problem to perform mission design, or to implement it on board spacecrafts due to its low computational requirements.

This work is organized as follows. First, an introduction of the problem and the hypothesis that are assumed is presented. Second, the formulation for an absolute station keeping of a satellite is described, which is the basis of the formulation that is later used in the tandem formation study. Third, the model for tandem formation is shown, which includes a simple set of equations to study the dynamic of the system. Fourth, the model presented is applied for the case of a master-slave scenario, which includes the study on the evolution of the system, its possible maneuvering strategies and the definition of a control law for the formation. Finally, an example of application is included to show the possibilities of this model.

%%%%%%%%%%%%%%%%%%%%%%%%%%%%%%%%%%%%%%%%%%%%%%%%%%%%%%%%%%%%%%%%%%%%%%%%%%%%%%%%%%%%%%%%%%%%%%
%%%%%%%%%%%%%%%%%%%%%%%%%%%%%%%%%%%%%%%%%%%%%%%%%%%%%%%%%%%%%%%%%%%%%%%%%%%%%%%%%%%%%%%%%%%%%%
%%%%%%%%%%%%%%%%%%%%%%%%%%%%%%%%%%%%%%%%%%%%%%%%%%%%%%%%%%%%%%%%%%%%%%%%%%%%%%%%%%%%%%%%%%%%%%

\section{Preliminaries}

Throughout this work we make use of the classical orbital elements, namely the semi-major axis ($a$), the inclination ($i$), the eccentricity ($e$), the argument of perigee $\omega$, the right ascension of the ascending node ($\Omega$) and the mean anomaly ($M$). Other important parameters used are the Earth gravitational constant ($\mu$), the term $J_2$ of the Earth gravitational potential (which is related to the oblateness of the Earth), the Earth spin rate $(\omega_{\oplus})$ and the Earth equatorial radius ($R_{\oplus}$).

%%%%%%%%%%%%%%%%%%%%%%%%%%%%%%%%%%%%%%%%%%%%%%%%%%%%%%%%%%%%%%%%%%%%%%%%%%%%%%%%%%%%%%%%%%%%%%
%%%%%%%%%%%%%%%%%%%%%%%%%%%%%%%%%%%%%%%%%%%%%%%%%%%%%%%%%%%%%%%%%%%%%%%%%%%%%%%%%%%%%%%%%%%%%%

\subsection{Hypothesis of the model}
\label{sec:hypothesis}

In order to develop a simple model which can be used in the study of the tandem formation between two or more satellites, some assumptions must be introduced. Thus, we present in this section all the hypothesis that are assumed during this work as a first order approximation to the problem.

First, the model introduced focuses on the study of satellites in near circular Low Earth Orbits (LEO). This means that, in general, the accelerations produced by the Earth gravitational potential and the atmospheric drag are the dominant perturbations, being other perturbations, such as the Sun and Moon as third bodies, the solar radiation pressure or the albedo, negligible when compared to the dominant perturbations.

Second, it is assumed that the nominal orbits of the satellites are defined by taking into account of effects of the Earth gravitational potential. This means that these nominal orbits are in fact perturbed trajectories under the perturbation produced by the Earth gravitational potential. For instance, and for the case of repeating ground-track orbits, these nominal orbits correspond to the orbits that, under the effects of just the Earth gravitational potential, maintain the repeating ground-track property without requiring additional orbital maneuvers~\cite{Vallado,wagner,tesis}. The objective of the definition of these nominal orbits is to allow the decoupling of the effects of the atmospheric drag and the Earth gravitational potential. This means that later, during the study, we will be able to focus on the effects of the atmospheric drag. 

Third, a constant reference density is considered during the propagation of the system instead of a density that varies over time. This reference density is defined based on a prediction for the next maneuvering cycle, and thus, it presents a high uncertainty. However, even with this limitation, the model is able to provide the general behavior of the system, being the approximation more precise the better the prediction performed. Additionally, this model can be also used in its differential formulation to take into account a varying density over the dynamic of the system.

Fourth, all the orbital maneuvers required for the maintenance of the formation are assumed to be in plane impulses that are tangent to the orbit. In particular, each complete orbital maneuver will consist on two burn impulses via a Hohmann transfer. These maneuvers have the purpose to raise the orbit in such a way that satellites always lay inside the set of boundaries defined by their mission requirements~\cite{Vallado}.

%%%%%%%%%%%%%%%%%%%%%%%%%%%%%%%%%%%%%%%%%%%%%%%%%%%%%%%%%%%%%%%%%%%%%%%%%%%%%%%%%%%%%%%%%%%%%%
%%%%%%%%%%%%%%%%%%%%%%%%%%%%%%%%%%%%%%%%%%%%%%%%%%%%%%%%%%%%%%%%%%%%%%%%%%%%%%%%%%%%%%%%%%%%%%
%%%%%%%%%%%%%%%%%%%%%%%%%%%%%%%%%%%%%%%%%%%%%%%%%%%%%%%%%%%%%%%%%%%%%%%%%%%%%%%%%%%%%%%%%%%%%%

\section{Absolute maintenance}

In this section we propose a model to study the station keeping maneuvers required to maintain a satellite inside a control box around its nominal orbit. This model is defined by selecting the nominal orbit of the satellite as the reference for the dynamic, where the position of the perturbed satellite is described using the along track and cross track distances with respect to its nominal orbit. To that end, we first study the dynamic of a satellite subjected to the perturbation produced by the atmospheric drag, focusing on the evolution of its semi-major axis over time. Then, the station keeping, with defined boundaries both in the along track and cross track distances, is studied. Comparison with the effects of the term $J_2$ from the Earth gravitational potential are also included since it is the most important perturbation for LEO.

%%%%%%%%%%%%%%%%%%%%%%%%%%%%%%%%%%%%%%%%%%%%%%%%%%%%%%%%%%%%%%%%%%%%%%%%%%%%%%%%%%%%%%%%%%%%%%
%%%%%%%%%%%%%%%%%%%%%%%%%%%%%%%%%%%%%%%%%%%%%%%%%%%%%%%%%%%%%%%%%%%%%%%%%%%%%%%%%%%%%%%%%%%%%%

\subsection{Dynamic of the satellite}

In this first subsection, we deal with the general dynamic of the satellite under the effects of the atmospheric drag. In particular, we derive a set of equations to easily compute the orbit decay and the variation that the orbital period experiences under the effects of this perturbation. These results are the basis of the model introduced later. 

%%%%%%%%%%%%%%%%%%%%%%%%%%%%%%%%%%%%%%%%%%%%%%%%%%%%%%%%%%%%%%%%%%%%%%%%%%%%%%%%%%%%%%%%%%%%%%

\subsubsection{Orbit decay}
The derivative of the semi-major axis can be obtained from Gauss equations particularized for the case of circular orbits~\cite{Vallado}:
\begin{equation}
\displaystyle\frac{da}{dt} = 2\frac{a}{v}\gamma_{\theta},
\end{equation}
where $v$ is the velocity of the satellite and $\gamma_{\theta}$ is the perturbing acceleration in the direction of the movement. In the case of atmospheric drag, this acceleration can be approximated by:
\begin{equation}
\gamma_{\theta} = -\displaystyle\frac{1}{2}\rho\frac{S}{m}c_dv^2,
\end{equation}
where $\rho$ is the atmospheric density in the satellite position, $S$ is the cross section area, $m$ is the mass of the satellite, and $c_d$ is the drag coefficient of the satellite. On the other hand, from the equation of energy applied to the orbit, we can obtain the following expression:
\begin{equation}
\displaystyle\frac{v^2}{2} - \displaystyle\frac{\mu}{a} = -\displaystyle\frac{\mu}{2a},
\end{equation}
from where it is possible to derive the velocity of the satellite in the nominal orbit:
\begin{equation}
v = \sqrt{\displaystyle\frac{\mu}{a}}.
\end{equation}
The value of the velocity ($v$) derived is then used in order to obtain the derivative of the semi-major axis:
\begin{equation} \label{da}
\displaystyle\frac{da}{dt} = -\rho \displaystyle\frac{S}{m}c_d \sqrt{\mu a},
\end{equation}
which provides a very simple expression that allows to compute the dynamic of a satellite in a near circular orbit.

From Equation~\eqref{da}, and under the assumption that the atmospheric density is constant (see Section~\ref{sec:hypothesis}), a direct integration can be performed in order to obtain the variation that the semi-major axis has experienced ($\delta a$) in a given time ($t$):
\begin{equation} \label{delta_a}
\delta a = a - \left[\sqrt{a} - \displaystyle\frac{1}{2}\rho\frac{S}{m}c_d\sqrt{\mu}t\right]^2.
\end{equation}
Alternatively, and if a more simplified expression is required, a linear evolution on the semi-major axis can be assumed since the variations that this variable experiences are very small in time considered between orbital maneuvers. That way, by integrating Equation~\eqref{da}, the variation that the semi-major axis experiences over time can be obtained:
\begin{equation}\label{eq:dalinear}
\delta a = - \rho\displaystyle\frac{S}{m}c_d\sqrt{\mu a}t.
\end{equation}
Another possible approach is to consider the semi-major axis of the orbit near constant during the integration (since its variation is in general small when compared with its absolute value) while taking into account the variation of the atmospheric density. That way, Equation~\eqref{da} can be integrated to obtain:
\begin{equation}
\delta a = - \displaystyle\frac{S}{m}c_d\sqrt{\mu a}\int_0^t\rho dt.
\end{equation}

%%%%%%%%%%%%%%%%%%%%%%%%%%%%%%%%%%%%%%%%%%%%%%%%%%%%%%%%%%%%%%%%%%%%%%%%%%%%%%%%%%%%%%%%%%%%%%

\subsubsection{Along track drift}

Let $a_0$ be the nominal semi-major axis of the satellite orbit, that is, the semi-major axis that allows the achievement of the ground-track property of the orbit under the effects of the Earth gravitational potential. As a first order approximation to the problem, only the $J_2$ term of the Earth gravitational potential is taken into account. Therefore, the mean motion associated with the nominal orbit ($n_0$) can be obtained using the following expression~\cite{Vallado}:
\begin{equation} \label{period0j2}
n_0 = \sqrt{\displaystyle\frac{\mu}{a_0^3}}\left[1+\frac{3}{2}\frac{J_2R_{\oplus}}{a_0^2}(1-\frac{3}{2}\sin^2(i))(1-e^2)^{-3/2}\right],
\end{equation}
where $R_{\oplus}$ is the Earth radius at the equator. On the other hand, let $a$ be the actual mean semi-major axis of the orbit in a given instant. The value of $a$ can be expressed by means of the nominal semi-major axis as:
\begin{equation}
a = a_0 + \Delta a,
\end{equation} 
with $\Delta a$ being, in general, small compared to $a_0$. This means that the mean motion of the satellite for a given instant can be written as:
\begin{equation} \label{perioddeltaj2}
n = \sqrt{\displaystyle\frac{\mu}{(a_0 + \Delta a)^3}}\left[1+\frac{3}{2}\frac{J_2R_{\oplus}}{(a_0 + \Delta a)^2}(1-\frac{3}{2}\sin^2(i))(1-e^2)^{-3/2}\right].
\end{equation}
From the nominal and instantaneous mean motions of the satellite (Equations~\eqref{period0j2} and~\eqref{perioddeltaj2} respectively), it is possible to compute the time drift experienced by the satellite in an orbital period ($\Delta t_T$) with respect to its nominal orbit:
\begin{equation}
\Delta t_T = T_0 - T = 2\pi\left[\displaystyle\frac{1}{n_0} - \frac{1}{n}\right]
\end{equation}
or in a differential formulation:
\begin{equation} \label{deltatt}
\displaystyle\frac{d\Delta t}{dt} = n_0\left(\frac{1}{n_0} - \frac{1}{n}\right).
\end{equation}
Once this is done, a first order approximation is performed by a Taylor series expansion in the non-dimensional parameter $\Delta a/a$. That way, the inverse of the mean motion can be approximated by:
\begin{equation} \label{eq:inversen}
\displaystyle\frac{1}{n} \approx \frac{1}{n_0} - \frac{1}{n_0^2}\left[-\frac{3}{2}\sqrt{\displaystyle\frac{\mu}{a_0^3}} - \frac{21}{4}\frac{J_2R_{\oplus}\sqrt{\mu}}{a_0^{7/2}}(1-\frac{3}{2}\sin^2(i))(1-e^2)^{-3/2} \right]\frac{\Delta a}{a_0}.
\end{equation} 
Now, a comparison between the first and second order terms in the previous expression is performed, where it is easy to derive that:
\begin{equation}
\displaystyle\frac{\frac{3}{2}\sqrt{\displaystyle\frac{\mu}{a_0^3}}}{\frac{21}{4}\frac{J_2R_{\oplus}\sqrt{\mu}}{a_0^{7/2}}(1-\frac{3}{2}\sin^2(i))(1-e^2)^{-3/2}} \sim \frac{a_0^2}{J_2R_{\oplus}^2} \gg 1,
\end{equation}
which allows to obtain a first order approximation of Equation~\eqref{eq:inversen}:
\begin{equation}\label{periodj2}
\displaystyle\frac{1}{n} \approx \frac{1}{n_0} + \frac{3}{2}\frac{1}{n_0^2}\sqrt{\displaystyle\frac{\mu}{a_0^3}}\frac{\Delta a}{a_0}.
\end{equation}
Then, Equation~\eqref{periodj2} is introduced in Equation~\eqref{deltatt} and after some simple equation manipulations, the following expression is obtained:
\begin{equation}
\displaystyle\frac{d\Delta t}{dt} = -\frac{3}{2}\frac{1}{n_0}\sqrt{\displaystyle\frac{\mu}{a_0^3}}\frac{\Delta a}{a_0}.
\end{equation}
Additionally, $n_0$ can be approximated by:
\begin{equation}\label{n0}
n_0 \approx \sqrt{\displaystyle\frac{\mu}{a_0^3}},
\end{equation}
since the term in $J_2$ is a thousand times smaller that the one that is maintained. That way, the derivative of the time drift is:
\begin{equation}
\displaystyle\frac{d\Delta t}{dt} = -\frac{3}{2}\frac{\Delta a}{a_0}.
\end{equation}
and, if we assume that the orbit decay is linear during a small period of time (see Equation~\eqref{eq:dalinear}), we obtain:
\begin{equation} \label{eq:deltatgeneral}
\displaystyle\frac{d\Delta t}{dt} =  -\frac{3}{2}\frac{\Delta a_0 + \frac{da}{dt}t}{a_0},
\end{equation}
where $\Delta a_0$ is the initial semi-major axis of the satellite related to the nominal orbit, that is, $a = a_0 + \Delta a_0$. Note that Equation~\eqref{eq:deltatgeneral} is valid for orbits with any eccentricity and scenarios where the density in non-constant since no simplification has been done during the process in that respect. If instead we focus on the case of circular orbits, the previous expression can be rewritten as:
\begin{equation}
\displaystyle\frac{d\Delta t}{dt} =  -\frac{3}{2}\frac{\Delta a_0 - \rho\frac{S}{m}c_d\sqrt{\mu a_0}t}{a_0}.
\end{equation}
Then, if the density is considered to be constant, the integral can be calculated analytically, leading to:
\begin{equation}\label{deltatnominal}
\Delta t = \Delta t_0 - \displaystyle\frac{3}{2}\frac{\Delta a_0}{a_0}t + \frac{3}{4}\rho\frac{S}{m}c_d\sqrt{\frac{\mu}{a_0}}t^2,
\end{equation}
which provides the evolution of the along track drift of a satellite with respect to its nominal orbit by the use of a very simple and compact expression. Note also that Equation~\eqref{deltatnominal} only depends on the initial conditions of the satellite ($\Delta t_0$, $\Delta a_0$ and $a_0$), its physical properties ($S$, $m$ and $c_d$), the atmospheric density ($\rho$) and the time in which the drift is evaluated ($t$). This means that it is possible to define the whole dynamic of the system by the definition of these parameters, no requiring further information.

%%%%%%%%%%%%%%%%%%%%%%%%%%%%%%%%%%%%%%%%%%%%%%%%%%%%%%%%%%%%%%%%%%%%%%%%%%%%%%%%%%%%%%%%%%%%%%
%%%%%%%%%%%%%%%%%%%%%%%%%%%%%%%%%%%%%%%%%%%%%%%%%%%%%%%%%%%%%%%%%%%%%%%%%%%%%%%%%%%%%%%%%%%%%%

\subsection{Cross track maintenance}

The objective now is to derive a series of equations that allow to compute in a simple and compact manner the maneuver frequency, size of the maneuver and impulse per maneuver that a satellite requires in order to maintain its orbit in a cross track boundary defined by its mission requirements. In that respect, it is assumed that, since the only orbit perturbation considered is the atmospheric drag, the maximum drift of the ground-tracks happens when satellites fly over at the Earth equator. If other perturbations were taken into account, such as the Sun as third body, this assumption would be no longer applicable, as the inclination can also experience variations due to these orbital perturbations.

%%%%%%%%%%%%%%%%%%%%%%%%%%%%%%%%%%%%%%%%%%%%%%%%%%%%%%%%%%%%%%%%%%%%%%%%%%%%%%%%%%%%%%%%%%%%%%

\subsubsection{Maneuver frequency}

In order to compute the maneuver frequency, it is first required to know the rate of change that the deviation of the ground-track experiences over time. Once this result is obtained, the evolution of the ground-track drift will be computed by the integration of its rate of change. Finally, a relation between the time between maneuvers (maneuver frequency) and the dead band requirement will be established using these expressions.

Let $\Delta \lambda$ be the angle shifted with respect to the nominal definition that the ground-track has experienced over the Equator at a given time $t$, being $\Delta\lambda$ defined as positive when the ground-track deviation is towards the East and negative when it is towards the West. Then, the ground-track drift at a given time can be obtained using the result from Equation~\eqref{deltatnominal} applied to the resultant angle shifted during the rotation of the Earth:
\begin{equation} \label{lambda}
\Delta\lambda = \Delta\lambda_0 -\displaystyle\frac{3}{2}\omega_{\oplus}\frac{\Delta a_0}{a_0}t + \frac{3}{4}\omega_{\oplus}\rho\frac{S}{m}c_d\sqrt{\frac{\mu}{a_0}}t^2
\end{equation}
where $\Delta\lambda_0$ is the ground-track drift when $t = 0$.

\begin{figure}[h!]
	\centering
	\includegraphics[width=0.6\textwidth]{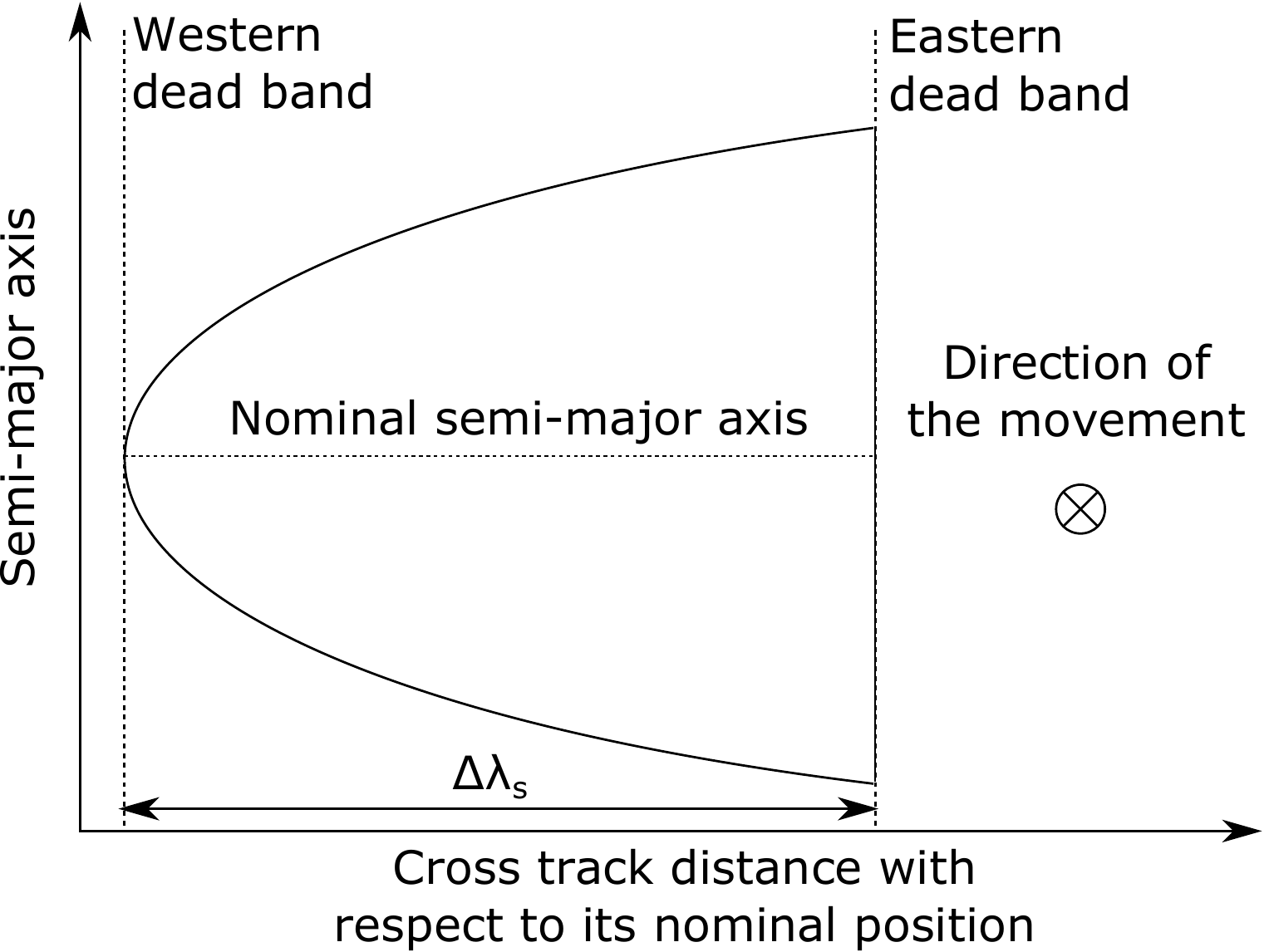} 
	\caption{Cross track radial dynamic of the satellite.}
	\label{fig:cross}
\end{figure}

Now, we are interested to know the time frequency between maneuvers. Let $\Delta\lambda_s$ be the total dead band size that the mission is aiming for, that is, the maximum angle allowed by the mission requirements measured from the western to the eastern boundary. From Equation~\eqref{lambda}, the derivative of the ground-track drift can be obtained:
\begin{equation} \label{dlambda}
\displaystyle\frac{d\Delta\lambda}{dt} = -\displaystyle\frac{3}{2}\omega_{\oplus}\frac{\Delta a_0}{a_0} + \frac{3}{2}\omega_{\oplus}\rho\frac{S}{m}c_d\sqrt{\frac{\mu}{a_0}}t,
\end{equation}
where it can be noted that the change in the direction of the drift happens when the semi-major axis of the orbit is equal to $a_0$, and thus, $\Delta a_0 = 0$. This also implies that, as the orbit decays over time, the point corresponding to the western boundary shall be the one where $a = a_0$ in order to impose the boundary conditions of the mission. Let this point be the starting condition for this study (see also Figure~\ref{fig:cross}). That way, it is possible to calculate half the maneuvering period ($T_{DB}$) using Equation~\eqref{lambda}:
\begin{equation}
T_{DB} = 2\sqrt{\frac{\Delta\lambda_s}{3\omega_{\oplus}\rho \displaystyle\frac{S}{m}c_d}\sqrt{\displaystyle\frac{a_0}{\mu}}}.
\end{equation}
Thus, the maneuvering frequency ($T_{M_{\text{cross}}}$) to compensate the atmospheric drag and maintain the dead band defined is:
\begin{equation}\label{tm}
T_{M_{\text{cross}}} = 4\sqrt{\frac{\Delta\lambda_s}{3\omega_{\oplus}\rho \displaystyle\frac{S}{m}c_d}\sqrt{\displaystyle\frac{a_0}{\mu}}}.
\end{equation}
It is important to note that once the requirements of the mission ($\Delta\lambda_s$), the satellite physical properties ($\frac{S}{m}c_d$), and the nominal semi-major axis of the orbit ($a_0$) are established, Equation~\eqref{tm} only depends on the atmospheric density at the altitude of the satellite. In particular:
\begin{equation}
T_{M_{\text{cross}}} = 4\sqrt{\frac{\Delta\lambda_s}{3\omega_{\oplus} \displaystyle\frac{S}{m}c_d}\sqrt{\displaystyle\frac{a_0}{\mu}}}\sqrt{\frac{1}{\rho}}, \quad \text{with} \quad 4\sqrt{\frac{\Delta\lambda_s}{3\omega_{\oplus} \displaystyle\frac{S}{m}c_d}\sqrt{\displaystyle\frac{a_0}{\mu}}} = \text{cte}.
\end{equation}

In the previous result, the model assumed that the orbital plane was coincident with the instantaneous orbital plane resultant from the propagation of the nominal orbit. This means that the considered orbital plane drifts due to the effect of the gravitational potential of the Earth at the same rate than the nominal orbit, but this also implies that this drift does not produce any relative movement in the ground-track since nominal orbits are assumed to perfectly close their ground-tracks under the effect of the Earth gravitational potential by definition. However, since the $J_2$ perturbation is the largest perturbing force affecting the problem in LEO, we still have to study the effects that the variation of the semi-major axis produce in the rate at which the orbital plane drifts, which could affect the relative moment of the ground-track of the satellite.

The $J_2$ perturbation produces a drift in the right ascension of the ascending node of the satellite orbits. This effect is bigger the closer is the satellite to the Earth. Let $\Delta\lambda_{\Omega}$ be the drift produced by this perturbation with respect to the nominal orbit in a given instant. Then, its dynamic can be defined as:
\begin{equation}
\Delta\lambda_{\Omega} = \int \left(\displaystyle\frac{d\Omega_0}{dt} - \displaystyle\frac{d\Omega}{dt}\right)dt,
\end{equation}
where $\Omega_0$ relates to the nominal orbit and $\Omega$ to the perturbed orbit. Using the secular variation of the derivative of $\Omega$ produced by the $J_2$ perturbation:
\begin{equation}
\displaystyle\frac{d\Omega}{dt} = -\frac{3J_2R_{\oplus}^2}{2a^2(1-e^2)^2}\sqrt{\displaystyle\frac{\mu}{a^3}}\cos(i),
\end{equation}
we can obtain that:
\begin{equation}
\Delta\lambda_{\Omega} = \int \left(-\displaystyle\frac{3J_2R_{\oplus}^2\sqrt{\mu}}{2(1-e^2)^2}a_0^{-7/2}\cos(i) + \displaystyle\frac{3J_2R_{\oplus}^2\sqrt{\mu}}{2(1-e^2)^2}(a_0 + \delta a)^{-7/2}\cos(i)\right)dt,
\end{equation}
where $\delta a \ll a_0$ is the variation of the semi-major axis of the orbit with respect to its nominal value. As before, a Taylor series expansion is performed in this expression using the variation of the semi-major axis $\delta a/a_0$ as the variable, obtaining:
\begin{eqnarray}
\Delta\lambda_{\Omega} & = & \int \left(-\displaystyle\frac{21}{4}J_2R_{\oplus}^2\frac{\sqrt{\mu}\cos(i)}{(1-e^2)^2}a_0^{-9/2}\delta a\right)dt =\nonumber\\
& = & \int \left(-\displaystyle\frac{21}{4}J_2R_{\oplus}^2\frac{\sqrt{\mu}\cos(i)}{(1-e^2)^2}a_0^{-9/2}\left(\Delta a_0 + \frac{da}{dt}t\right)\right)dt.
\end{eqnarray}
Finally, by performing the integration for the case of near circular orbits and constant density, the evolution of the drift is obtained:
\begin{equation} \label{lambdaraan}
\Delta\lambda_{\Omega} = \Delta\lambda_{0} - \displaystyle\frac{21}{8}J_2R_{\oplus}^2\sqrt{\mu}\cos(i)a_0^{-9/2}\left(2\Delta a_0 t - \rho\frac{S}{m}c_d\sqrt{\mu a_0}t^2\right).
\end{equation}
%\begin{equation} \label{lambdaraan}
%\Delta\lambda_{\Omega} = \Delta\lambda_{0} - \displaystyle\frac{21}{8}J_2R_{\oplus}^2\frac{\sqrt{\mu}\cos(i)}{2(1-e^2)^2}a_0^{-9/2}\left(2\Delta a_0 t + \frac{da}{dt}t^2\right).
%\end{equation}

The objective now is to compare the deviation due to the delay in the period with the one produced by the drift in the orbital plane. Without loss of generality, let $\Delta\lambda_0 = 0$. Then, by using Equations~\eqref{lambda} and~\eqref{lambdaraan}:
\begin{equation}
\displaystyle\frac{\Delta\lambda_{\Omega}}{\Delta\lambda} = \frac{7}{2}J_2\sqrt{\displaystyle\frac{\mu R_{\oplus}^2}{a_0^7}}\cos(i) \ll 1.
\end{equation}
This shows that, in general, the effect of the drift in the orbital plane is negligible compared with the effect of the variation in the orbital period of the orbit.

%%%%%%%%%%%%%%%%%%%%%%%%%%%%%%%%%%%%%%%%%%%%%%%%%%%%%%%%%%%%%%%%%%%%%%%%%%%%%%%%%%%%%%%%%%%%%%

\subsubsection{Size of the maneuvers}

Once the time between maneuvers has been calculated, it is now possible to compute the size of the maneuver. From Equations~\eqref{da} and~\eqref{tm}, the size of the maneuver is directly obtained:
\begin{equation}
\Delta a = \rho\displaystyle\frac{S}{m}c_d\sqrt{\mu a_0}T_{M_{\text{cross}}} = \frac{4}{\sqrt{3}}\sqrt{\frac{\Delta\lambda_s}{\omega_{\oplus}}\rho\displaystyle\frac{S}{m}c_d\sqrt{\mu a_0^3}},
\end{equation}
where $\rho$ is the reference density during the maneuvering cycle, and the rest of parameters only depend on the satellite and the nominal orbit.

%%%%%%%%%%%%%%%%%%%%%%%%%%%%%%%%%%%%%%%%%%%%%%%%%%%%%%%%%%%%%%%%%%%%%%%%%%%%%%%%%%%%%%%%%%%%%%

\subsubsection{Impulse required}

Having obtained the size of the maneuver, we can proceed deriving the expressions for the impulse required in each maneuver. In order to do that, we assume that the maneuver is based on a Hohmann transfer consisting on two impulses. The first impulse is performed in the original orbit in the direction of the movement in order to increase the velocity of the satellite and the semi-major axis in such a way that the apogee of the orbit is located in the final orbit. The second impulse is performed at the apogee of the transfer orbit in the opposite direction of the movement, and such that the final orbit presents the semi-major axis required.

Let $a_i$ and $a_f$ be the semi-major axes of the initial and final orbits respectively. Then, both semi-major axes can be defined in terms of the nominal orbit and the maneuver size, in particular:
\begin{equation}
a_i = a_0 - \displaystyle\frac{\Delta a}{2}, \qquad a_f = a_0 + \displaystyle\frac{\Delta a}{2}, \qquad 2a_0 = a_i + a_f.
\end{equation} 
If the initial and final orbits are circular orbits, the semi-major axis of the transfer orbit ($a_t$) is:
\begin{equation}
a_t = \displaystyle\frac{a_i+a_f}{2},
\end{equation}
and the speed of the satellite before and after the maneuvers are:
\begin{eqnarray}
v_i = \sqrt{\displaystyle\frac{\mu}{a_i}}, \qquad\qquad v_{ti} = \sqrt{\displaystyle\frac{2\mu}{a_i} - \frac{\mu}{a_t}}, \nonumber \\
v_f = \sqrt{\displaystyle\frac{\mu}{a_f}}, \qquad\qquad v_{tf} = \sqrt{\displaystyle\frac{2\mu}{a_f} - \frac{\mu}{a_t}},
\end{eqnarray}
where $v_i$ and $v_f$ are the velocity of the initial and final orbits; and $v_{ti}$ and $v_{tf}$ are the initial and final velocities of the satellite in the transfer orbit. Thus, the total impulse of the maneuvers ($\Delta v$) is computed through:
\begin{equation}
\Delta v = (v_{ti} - v_i) + (v_f - v_{tf}),
\end{equation}
which using the former expressions leads to:
\begin{equation} \label{dv}
\Delta v = \sqrt{\displaystyle\frac{\mu}{a_0}}\left[\sqrt{\displaystyle\frac{1 + \frac{\Delta a}{2a_0}}{1 - \frac{\Delta a}{2a_0}}} - \sqrt{\displaystyle\frac{1}{1 - \frac{\Delta a}{2a_0}}} + \sqrt{\displaystyle\frac{1}{1 + \frac{\Delta a}{2a_0}}} - \sqrt{\displaystyle\frac{1 - \frac{\Delta a}{2a_0}}{1 + \frac{\Delta a}{2a_0}}}\right].
\end{equation}
Equation~\eqref{dv} can be simplified by performing a first order Taylor series expansion in $\frac{\Delta a}{2a_0}$. That way:
\begin{equation} \label{dvs}
\Delta v = \sqrt{\displaystyle\frac{\mu}{a_0}}\frac{\Delta a}{2a_0} = \frac{2}{\sqrt{3}}\displaystyle\sqrt{\frac{\Delta\lambda_s}{\omega_{\oplus}}\rho\frac{S}{m}c_d\sqrt{\frac{\mu^3}{a_0^3}}}.,
\end{equation}
which is a much simpler expression to compute the impulse required in each maneuver.

%%%%%%%%%%%%%%%%%%%%%%%%%%%%%%%%%%%%%%%%%%%%%%%%%%%%%%%%%%%%%%%%%%%%%%%%%%%%%%%%%%%%%%%%%%%%%%
%%%%%%%%%%%%%%%%%%%%%%%%%%%%%%%%%%%%%%%%%%%%%%%%%%%%%%%%%%%%%%%%%%%%%%%%%%%%%%%%%%%%%%%%%%%%%%

\subsection{Along track maintenance}

In this section, the study of the along track maintenance of an orbit is presented. This provides a useful alternative to the definition of the mission requirements using the dead band of the ground-tracks that can have other applications such as imposing a boundary in the date of the orbits. In that respect, we focus in here on the computation of the frequency between in plane maneuvers, as well as on the size and impulse required for each maneuvering cycle.

%%%%%%%%%%%%%%%%%%%%%%%%%%%%%%%%%%%%%%%%%%%%%%%%%%%%%%%%%%%%%%%%%%%%%%%%%%%%%%%%%%%%%%%%%%%%%%

\subsubsection{Maneuver frequency}

\begin{figure}[h!]
	\centering
	\includegraphics[width=0.60\textwidth]{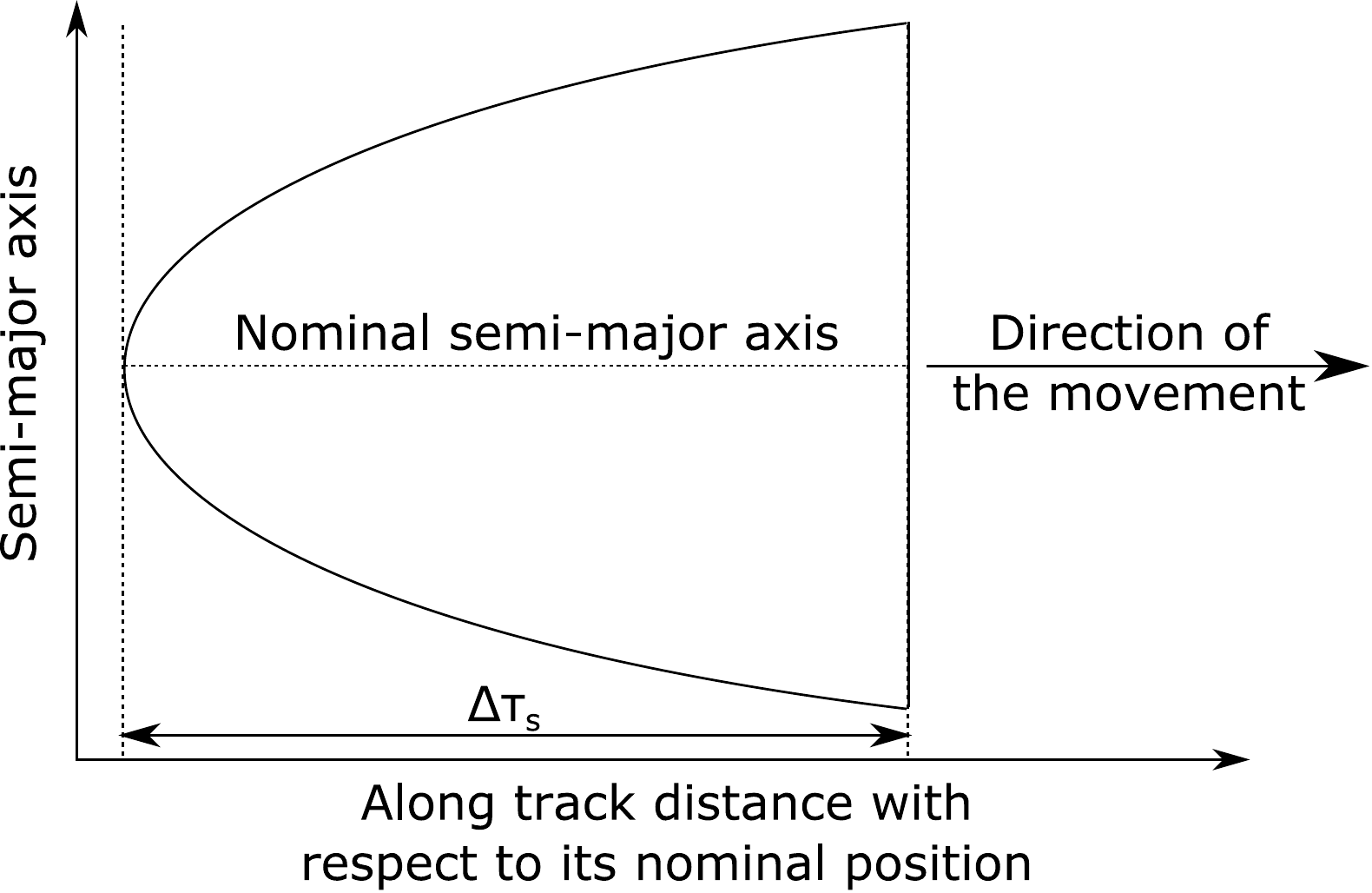} 
	\caption{Along track radial dynamic of the satellite.}
	\label{fig:along}
\end{figure}

Let $\Delta\tau_s$ be the range in  the along track time distance with respect to the reference orbit that the satellite is allowed to present. Let $\Delta a_0 = 0$ be the initial condition of the variation in the semi-major axis. Then, using Equation~\eqref{deltatnominal}:
\begin{equation}
T_b = 2\sqrt{\displaystyle\frac{-a_0\Delta\tau_s}{3\frac{da}{dt}}},
\end{equation}
where $T_b$ represents the time that the satellite takes to move from one boundary to the opposite (see Figure~\ref{fig:along}). Thus, the time between maneuvers is two times this value, as the opposite movement requires also to be taken into account. That way, the maneuver frequency ($T_M$) is:
\begin{equation} \label{tm2}
T_{M_{\text{along}}} = 4\sqrt{\displaystyle\frac{\Delta\tau_s}{3\rho\frac{S}{m}c_d}\sqrt{\frac{a_0}{\mu}}}.
\end{equation}

%%%%%%%%%%%%%%%%%%%%%%%%%%%%%%%%%%%%%%%%%%%%%%%%%%%%%%%%%%%%%%%%%%%%%%%%%%%%%%%%%%%%%%%%%%%%%%

\subsubsection{Size of the maneuver and impulse required}

The size of the maneuver and the impulse required follow the same equations shown for the case of cross track maintenance (Equations~\eqref{delta_a} and~\eqref{dvs}) but for the different maneuver frequency. Therefore, we do not repeat their derivation in here. Using Equation~\eqref{tm2}, the size of the maneuver is:
\begin{equation}
\Delta a_{\text{along}} = \rho\displaystyle\frac{S}{m}c_d\sqrt{\mu a_0}T_{M_{\text{along}}} = \frac{4}{\sqrt{3}}\sqrt{\Delta\tau_s\rho\displaystyle\frac{S}{m}c_d\sqrt{\mu a_0^3}},
\end{equation}
while the impulse required per maneuver is:
\begin{equation}
\Delta v_{\text{along}} = \sqrt{\displaystyle\frac{\mu}{a_0}}\frac{\Delta a_{\text{along}}}{2a_0}=  \frac{2}{\sqrt{3}}\displaystyle\sqrt{\Delta\tau_s\rho\frac{S}{m}c_d\sqrt{\frac{\mu^3}{a_0^3}}}.
\end{equation}

%%%%%%%%%%%%%%%%%%%%%%%%%%%%%%%%%%%%%%%%%%%%%%%%%%%%%%%%%%%%%%%%%%%%%%%%%%%%%%%%%%%%%%%%%%%%%%
%%%%%%%%%%%%%%%%%%%%%%%%%%%%%%%%%%%%%%%%%%%%%%%%%%%%%%%%%%%%%%%%%%%%%%%%%%%%%%%%%%%%%%%%%%%%%%

\subsection{Along and cross boundary relation}

In the former sections, the along and cross maintenance of the ground-track of the orbit were considered. However, each effect was studied separately. The objective now is to define a relation between both effects in order to determine the most restrictive condition for the mission.

From Equations~\eqref{tm} and~\eqref{tm2} a relation between both maneuver frequencies can be obtained:
\begin{equation}
\displaystyle\frac{T_{M_{\text{cross}}}}{T_{M_{\text{along}}}} = \sqrt{\displaystyle\frac{\Delta\lambda_s}{\omega_{\oplus}\Delta\tau_s}},
\end{equation}
where it can be clearly seen that it only depends on the square root of the relation between the size of the dead band and the available range in the along track time distance.

%%%%%%%%%%%%%%%%%%%%%%%%%%%%%%%%%%%%%%%%%%%%%%%%%%%%%%%%%%%%%%%%%%%%%%%%%%%%%%%%%%%%%%%%%%%%%%
%%%%%%%%%%%%%%%%%%%%%%%%%%%%%%%%%%%%%%%%%%%%%%%%%%%%%%%%%%%%%%%%%%%%%%%%%%%%%%%%%%%%%%%%%%%%%%
%%%%%%%%%%%%%%%%%%%%%%%%%%%%%%%%%%%%%%%%%%%%%%%%%%%%%%%%%%%%%%%%%%%%%%%%%%%%%%%%%%%%%%%%%%%%%%

\section{Tandem formation}

In this section, a study of the tandem formation of two satellites laying in the same nominal ground-track is presented. This means that both satellites will observe the same regions of the Earth under the same geometry, a very important property in many Earth observation missions. In addition, a master-slave scenario is considered, where the master satellite will perform its operations as if not in formation, while the slave will be subjected to the dynamic of its master. This can represent a scenario where one satellite of the formation acts as a leader, or a mission where a satellite is required to work in cooperation with another satellite already in orbit.

In order to perform this study, several considerations must be taken into account. First, both satellites can have different values of their ballistic coefficients. This means that the considered satellites can present different designs, or alternatively, that each satellite is subjected to different conditions. Second, there should be a delay between the moment in which the second satellite performs its maneuver and the first one. This condition represents the fact that, due to the process required for the maneuvering operations, some time is needed between the maneuver of the first satellite and the second in order to assure the identification of the final orbit from the first satellite, and to define the proper impulse for the second satellite.

That way, and as a result of the present study, it can be derived that the dynamic of the system will depend primary on the following parameters: the control strategy selected, the atmospheric density, the ratio between the ballistic coefficients of both satellites and the time delay between the maneuvers of the satellites. In that regard, we introduce two maneuvering strategies in this work, which are then studied as a function of the rest of the mentioned parameters.

%%%%%%%%%%%%%%%%%%%%%%%%%%%%%%%%%%%%%%%%%%%%%%%%%%%%%%%%%%%%%%%%%%%%%%%%%%%%%%%%%%%%%%%%%%%%%%
%%%%%%%%%%%%%%%%%%%%%%%%%%%%%%%%%%%%%%%%%%%%%%%%%%%%%%%%%%%%%%%%%%%%%%%%%%%%%%%%%%%%%%%%%%%%%%

\subsection{Nominal definition of the system}

As the tandem formation requires to share the same nominal ground-track, that is, all the satellites follow the same relative space-track from the Earth Centered Earth Fixed (ECEF) frame of reference, we use a formulation based on the along track time distances between satellites to define this particular configuration. In particular, in Arnas et al.~\cite{Time,Time2}, the definition of the constellation is performed directly in the ECEF frame of reference, being able to define a set of ground-tracks where the satellites of the constellation are located. That way, and following this formulation, the relative distribution of the constellation can be defined as:
\begin{eqnarray}
\Delta \Omega_{kq} & = & - \omega_{\oplus}t_{q}, \nonumber \\
\Delta M_{kq} & = & n\left(t_k + t_q\right);
\label{timeconstellations}
\end{eqnarray} 
where $\Delta \Omega_{kq}$, $\Delta M_{kq}$ are the values of the right ascension of the ascending node and the mean anomaly respect to the reference satellite respectively. On the other hand, $t_q$ and $t_k$ represent the distribution in time of the satellites along the different relative trajectories, in such a way that satellite $(k, \: q)$ is placed in the position $q$ of the relative trajectory $k$. All other orbital elements (semi-major axis, eccentricity, inclination and argument of perigee) are common for all the satellites of the constellation. As we are interested in a tandem formation, only one relative trajectory is required ($t_k = 0$), and thus:
\begin{eqnarray}
\Delta \Omega_{q} & = & - \omega_{\oplus}t_{q}, \nonumber \\
\Delta M_{q} & = & nt_q;
\end{eqnarray}
being $t_q$ a distribution parameters representing the time that it takes for one satellite to pass over the same position in the ECEF frame of reference after the reference satellite. This distribution means that, for satellites flying in tandem, each satellite presents a different orbit, or more precisely, that all satellites from the tandem have a slightly rotated orbital plane with respect to the reference orbit.

%%%%%%%%%%%%%%%%%%%%%%%%%%%%%%%%%%%%%%%%%%%%%%%%%%%%%%%%%%%%%%%%%%%%%%%%%%%%%%%%%%%%%%%%%%%%%%
%%%%%%%%%%%%%%%%%%%%%%%%%%%%%%%%%%%%%%%%%%%%%%%%%%%%%%%%%%%%%%%%%%%%%%%%%%%%%%%%%%%%%%%%%%%%%%

\subsection{Evolution of the time distance between two satellites}

Let $bc_r$ be the ballistic coefficient ratio between two satellites, that is:
\begin{equation}
bc_r = \displaystyle\frac{bc_2}{bc_1},
\end{equation}
where
\begin{equation}
bc_1 = \displaystyle\frac{m_1}{S_1c_{d_1}}, \qquad \text{and} \qquad bc_2 = \displaystyle\frac{m_2}{S_2c_{d_2}},
\end{equation}
are the ballistic coefficients of the master satellite and the slave satellite respectively, and $m$, $S$ and $c_d$ are the mass, the cross surface and the drag coefficient of each satellite. Thus, the along track deviation of each satellite with respect to its nominal definition can be expressed by means of the dynamic of the master satellite. By using Equation~\eqref{deltatnominal}:
\begin{eqnarray}
\Delta t_1 & = & \Delta t_{10} - \displaystyle\frac{3}{2}\frac{\Delta a_{10}}{a_0}t + \frac{3}{4}\frac{\rho}{br_1}\sqrt{\frac{\mu}{a_0}}t^2, \nonumber \\
\Delta t_2 & = & \Delta t_{20} - \displaystyle\frac{3}{2}\frac{\Delta a_{20}}{a_0}t + \frac{3}{4}\frac{\rho}{br_1}\frac{1}{bc_r}\sqrt{\frac{\mu}{a_0}}t^2,
\end{eqnarray}
being $\Delta t_1$ and $\Delta t_2$ the along track distances of each satellite with respect to their nominal orbits,  $\Delta t_{10}$ and $\Delta t_{20}$ are their initial along track distances with respect to the nominal, and $\Delta a_{10}$ and $\Delta a_{20}$ are the initial semi-major axes of both satellites with respect to the reference orbit.

As both satellites are defined in the same nominal trajectory, it is possible to relate the along track position of one satellite to the other. In particular, the along track distance between both satellites can be expressed as:
\begin{equation}\label{dt}
\Delta t = \Delta t_{0} - \displaystyle\frac{3}{2}\frac{\Delta a_{20} - \Delta a_{10}}{a_0}t + \frac{3}{4}\frac{\rho}{bc_1}\sqrt{\frac{\mu}{a_0}}\left(\frac{1}{bc_r} - 1\right)t^2,
\end{equation}
where:
\begin{equation}
\Delta t_{0} = \Delta t_{20} - \Delta t_{10},
\end{equation}
is the initial along track distance between both satellites. As it can be seen in Equation~\eqref{dt}, the dynamic only depends on the density, the ballistic coefficient, and the initial position of the slave satellite with respect to its master (represented by the quantities $\Delta t_{0}$ and $\Delta a_{20} - \Delta a_{10}$). Moreover, it is also possible to determine the differential form of Equation~\eqref{dt}. From Equation~\eqref{eq:deltatgeneral}, we obtain:
\begin{equation}\label{eq:diftandem}
\displaystyle\frac{d\Delta t}{dt} = -\frac{3}{2}\frac{\Delta a_{20} - \Delta a_{10}}{a_0} - \frac{3}{2}\frac{\rho}{bc_1}\sqrt{\frac{\mu}{a_0}}\left(\frac{1}{bc_r} - 1\right)t,
\end{equation}
which can be used to perform an integration taking into account a non-constant density during the dynamic.

%%%%%%%%%%%%%%%%%%%%%%%%%%%%%%%%%%%%%%%%%%%%%%%%%%%%%%%%%%%%%%%%%%%%%%%%%%%%%%%%%%%%%%%%%%%%%%

\subsubsection{Maximum variation of the along track distance}

The extreme of the time distance between two satellites can be easily obtained using Equation~\eqref{dt} by calculating its derivative:
\begin{equation}
\displaystyle\frac{d\Delta t}{dt} = -\frac{3}{2}\frac{\Delta a_{20} - \Delta a_{10}}{a_0} - \frac{3}{2}\frac{\rho}{bc_1}\sqrt{\frac{\mu}{a_0}}\left(\frac{1}{bc_r} - 1\right)t = 0.
\end{equation}
This expression allows to derive that the time in which the extreme ($t_{extreme}$) happens is:
\begin{equation}
t_{extreme} = \displaystyle\frac{\Delta a_{20} - \Delta a_{10}}{\displaystyle\frac{\rho}{bc_1}\sqrt{\mu a}\left(\frac{1}{bc_r} - 1\right)},
\end{equation}
which can be introduced in Equation~\eqref{dt} in order to obtain the maximum variation of the time distance between two satellites:
\begin{equation}\label{max}
\Delta\left(\Delta t_{max}\right) = - \displaystyle\frac{3}{4}\frac{\left(\Delta a_{20} - \Delta a_{10}\right)^2}{a\displaystyle\frac{\rho}{bc_1}\sqrt{\mu a}\left(\frac{1}{bc_r} - 1\right)}.
\end{equation}
It is important to note that this extreme can represent the closest point or the farthest point in the dynamic, depending on the ratio of ballistic coefficients and the position of the slave satellite with respect to its master, that is, if the slave is ahead or behind its master.

%%%%%%%%%%%%%%%%%%%%%%%%%%%%%%%%%%%%%%%%%%%%%%%%%%%%%%%%%%%%%%%%%%%%%%%%%%%%%%%%%%%%%%%%%%%%%%
%%%%%%%%%%%%%%%%%%%%%%%%%%%%%%%%%%%%%%%%%%%%%%%%%%%%%%%%%%%%%%%%%%%%%%%%%%%%%%%%%%%%%%%%%%%%%%

\subsection{Basic tandem maneuvering strategy}

One of the most simple control strategies for a master-slave scenario mission consist of imposing the slave satellite to mimic all the in plane maneuvers that its master performs during its dynamic. This means that both satellites will share the same time frequency between maneuvers. However, as the ballistic coefficients of both satellites are different, the size of the maneuvers as well as the dead bands of both satellites are different.

\begin{figure}[h!]
	\centering
	\includegraphics[width=0.6\textwidth]{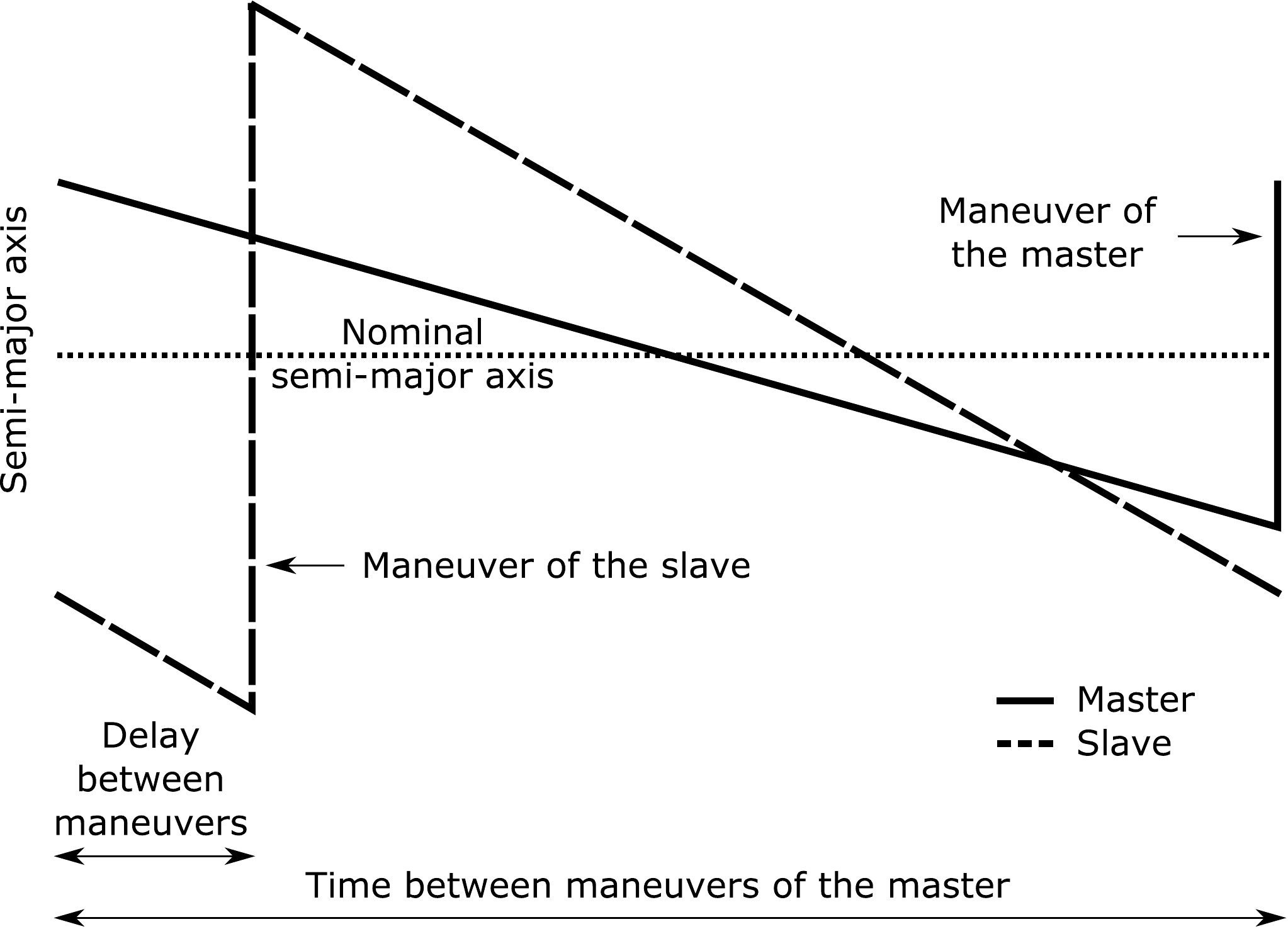} 
	\caption{Basic tandem maneuvering strategy concept.}
	\label{fig:concept1}
\end{figure}

Figure~\ref{fig:concept1} shows the graphical representation of the idea behind this maneuvering strategy. As it can be seen, both satellites have different variations in their semi-major axis due to different ballistic coefficients. Moreover, it can also be observed that there is a delay between the maneuvers of the master and the slave satellites, which changes slightly the dynamic of the system but maintains the frequency at which the maneuvers of both satellites are performed.

%%%%%%%%%%%%%%%%%%%%%%%%%%%%%%%%%%%%%%%%%%%%%%%%%%%%%%%%%%%%%%%%%%%%%%%%%%%%%%%%%%%%%%%%%%%%%%

\subsubsection{Relation in the maintenance boundaries}

The control that has been selected imposes that both satellites must have the same maneuvering frequency but for a delay between the maneuvers of the slave and master satellites. Let $T_M$ be the most restrictive maneuvering frequency from the along and the cross track maintenance of the master satellite. Then relating to the cross track deviation of both satellites:
\begin{equation}
T_M = 4\sqrt{\frac{\Delta\lambda_{s1}bc_1}{3\omega_{\oplus}\rho }\sqrt{\displaystyle\frac{a_0}{\mu}}} = 4\sqrt{\frac{\Delta\lambda_{s2}bc_1bc_r}{3\omega_{\oplus}\rho }\sqrt{\displaystyle\frac{a_0}{\mu}}}
\end{equation}
where $\Delta\lambda_{s1}$ and $\Delta\lambda_{s2}$ are the dead bands of the master and the slave satellite respectively. By performing some basic operations, a relation between both dead bands can be obtained:
\begin{equation}
\Delta\lambda_{s2} = \displaystyle\frac{\Delta\lambda_{s1}}{bc_r},
\end{equation}
which means that if the ballistic coefficient of the slave is smaller, its dead band will be larger and vice versa.

On the other hand, if we relate the frequency between maneuvers with the along track deviation of both satellites, this relation is obtained:
\begin{equation}
T_{M} = 4\sqrt{\displaystyle\frac{\Delta\tau_{s1}bc_1}{3\rho}\sqrt{\frac{a_0}{\mu}}} = 4\sqrt{\displaystyle\frac{\Delta\tau_{s2}bc_1bc_r}{3\rho}\sqrt{\frac{a_0}{\mu}}},
\end{equation}
where $\Delta\tau_{s1}$ and $\Delta\tau_{s2}$ are the boundaries in along track distance of the master and the slave respectively. This expression can be used to relate the boundaries of both satellites with respect to their nominal orbits:
\begin{equation}
\Delta\tau_{s2} = \displaystyle\frac{\Delta\tau_{s1}}{bc_r}.
\end{equation}

Finally, the size of the maneuvers of the master and slave satellites is compared. As both present the same frequency between maneuvers, the size of their maneuvers is:
\begin{eqnarray}
\Delta a_1 & = & \displaystyle\frac{\rho}{bc_1}\sqrt{\mu a_0}T_M, \nonumber \\
\Delta a_2 & = & \displaystyle\frac{\rho}{bc_1bc_r}\sqrt{\mu a_0}T_M,
\end{eqnarray}
being $\Delta a_1$ and $\Delta a_2$ the maneuver size of the master and the slave respectively. Thus, the relation between both maneuver sizes is:
\begin{equation}
\Delta a_2 = \displaystyle\frac{\Delta a_1}{bc_r}.
\end{equation}
Therefore, the relation in the impulse required for both satellites in each impulsive maneuver is:
\begin{equation}
\Delta v_2 = \displaystyle\frac{\Delta v_1}{bc_r},
\end{equation}
where $\Delta v_1$ and $\Delta v_2$ are the impulses required in each maneuver for the master and the slave satellite respectively.

%%%%%%%%%%%%%%%%%%%%%%%%%%%%%%%%%%%%%%%%%%%%%%%%%%%%%%%%%%%%%%%%%%%%%%%%%%%%%%%%%%%%%%%%%%%%%%

\subsubsection{Nominal operation of the maneuvering strategy}

From now on we will assume that $T_M$ is the frequency between maneuvers of the most restrictive boundary for the master satellite. That way, by using Equation~\eqref{deltatnominal} and imposing a complete maneuvering cycle:
\begin{equation}
0 = -\displaystyle\frac{3}{2}\frac{\Delta a_{10}}{a_0}T_M + \frac{3}{4}\frac{\rho}{bc_1}\sqrt{\frac{\mu}{a_0}}T_M^2,
\end{equation}
the initial condition of the master satellite after its maneuver can be obtained:
\begin{equation}
\Delta a_{10} = \displaystyle\frac{\rho}{bc_1}\sqrt{\mu a_0}\frac{T_M}{2}.
\end{equation}
The same operation can be done for the slave satellite obtaining:
\begin{equation}
\Delta a_{20} = \displaystyle\frac{\rho}{bc_1br_r}\sqrt{\mu a_0}\frac{T_M}{2}.
\end{equation}
However, it is important to note that as the maneuver of both satellites is performed at different moments, we cannot apply this initial conditions directly in Equation~\eqref{dt}.

Let assume that the master satellite always performs its maneuvers first and that the slave satellite is flying ahead of its master. In that case, just after the slave satellite has performed its maneuver, the master has already decayed a quantity proportional to the time delay between maneuvers. In particular:
\begin{equation}
\Delta a_{10} = \displaystyle\frac{\rho}{bc_1}\sqrt{\mu a_0}\left(\frac{T_M}{2} - t_d\right),
\end{equation}
where $t_d$ is the time delay between the maneuvers of the master and the ones of the slave satellite. Now, we can introduce this result in Equation~\eqref{dt} to obtain the dynamic of the system for the time period ranging between the maneuver of the slave and the one of the master:
\begin{equation}
\Delta t = \Delta t_{0} - \displaystyle\frac{3}{4}\frac{\rho}{bc_1}\sqrt{\frac{\mu}{a_0}}\left[\left(\left(\frac{1}{bc_r} - 1\right)T_M + 2t_d\right)t - \left(\frac{1}{bc_r} - 1\right)t^2\right],
\end{equation}
where time $t$ is related to the end of the maneuver of the slave satellite, and $\Delta t_0$ is the maximum (if $bc_r < 1$, or minimum if $bc_r > 1$) along track distance of the satellites in the dynamic (which is also the initial along track distance for this period). On the other hand, for the time period ranging between the maneuver of the master and the one of the slave, the dynamic is governed by:
\begin{eqnarray}
\Delta t & = & \Delta t_{0} - \displaystyle\frac{3}{4}\frac{\rho}{bc_1}\sqrt{\frac{\mu}{a_0}}\Bigg[\left(\frac{1}{bc_r} - 1\right)\left(T_M - t_d\right)t_d + \nonumber \\
& + & \left(2t_d - \left(\frac{1}{bc_r} + 1\right)T_M\right)t - \left(\frac{1}{bc_r} - 1\right)t^2\Bigg],
\end{eqnarray}
being $t$ related this time to the moment in which the master satellite performs its maneuver. Moreover it is possible to identify:
\begin{equation}
\Delta t_{1} = \Delta t_{0} - \displaystyle\frac{3}{4}\frac{\rho}{bc_1}\sqrt{\frac{\mu}{a_0}}\Bigg[\left(\frac{1}{bc_r} - 1\right)\left(T_M - t_d\right)t_d\Bigg]
\end{equation} 
as the along track distance between the satellites at the time where the master satellite performs its maneuver.

In either case, the maximum variation of the along track distance happens in the period between the maneuver of the slave satellite and the one of its master. In particular, the extreme of the along track distance is located in:
\begin{equation}
t_{\text{extreme}} = \displaystyle\frac{T_M}{2} + \frac{t_d}{\left(\displaystyle\frac{1}{bc_r} - 1\right)}.
\end{equation}
Thus, it is easy to derive that if the time delay between maneuvers fulfills the condition
$t_d < (1-bc_r)T_M/2$, the maximum variation of the along track distance is equal to:
\begin{equation}
\Delta \left(\Delta t_{max}\right) = -\displaystyle\frac{3}{4}\frac{\rho}{bc_1}\sqrt{\frac{\mu}{a_0}}\frac{\left(\frac{T_M}{2}\left(\frac{1}{bc_r} - 1\right) + t_d\right)^2}{\left(\frac{1}{bc_r} - 1\right)},
\end{equation}
where, if we consider the cross track maintenance as the most restrictive control requirement for the master satellite, we obtain:
\begin{equation}
\Delta \left(\Delta t_{max}\right) = -\displaystyle\frac{\left[\sqrt{\frac{\Delta\lambda_{s1}}{\omega_{\oplus}}}\left(\frac{1}{bc_r} - 1\right) + t_d\sqrt{\frac{3}{4}\frac{\rho}{bc_1}\sqrt{\frac{\mu}{a_0}}}\right]^2}{\left(\frac{1}{bc_r} - 1\right)}.
\end{equation}
Conversely, if $t_d > (1-bc_r)T_M/2$, the maximum variation happens just before the maneuver of the master satellite. That way, by using Equation~\eqref{dt} it is possible to obtain the maximum variation of the along track time distance between satellites:
\begin{equation}
\Delta \left(\Delta t_{max}\right) = -\displaystyle\frac{3}{4}\frac{\rho}{bc_1}\sqrt{\frac{\mu}{a_0}}\left(\frac{1}{bc_r} - 1\right)\left(T_M - t_d\right)t_d.
\end{equation}

Note that the case studied in here represents the situation where the master satellite always performs its maneuvers earlier, and the slave satellite flies ahead of its master. Other cases of study can be solved similarly to the one presented, being the only differences the changes in the initial conditions that are selected in order to apply Equation~\eqref{dt}.

%%%%%%%%%%%%%%%%%%%%%%%%%%%%%%%%%%%%%%%%%%%%%%%%%%%%%%%%%%%%%%%%%%%%%%%%%%%%%%%%%%%%%%%%%%%%%%

\subsubsection{Definition of control laws}

The model presented in the previous sections allows to easily define control laws for the system due to the simplicity of the model. Therefore, in this section we propose a very easy control devised to maintain satellites in their defined orbital boundaries, while at the same time, we take into account variations in the expected results for the impulse maneuvers and the atmospheric density during the dynamic of the system. Note that the control law presented in here is just an example of what can be done using the proposed model.

As the problem in study is based on a master-slave scenario, we assume that the control law of the master satellite is given by its mission, and thus, there is no possibility to interfere with it. This means that the control law proposed in this section will only apply to the slave satellite, having to adapt to the dynamic of its master.

Let $\Delta a_{10}$ be the planned initial position of the master satellite in semi-major axis with respect to its nominal orbit, and let $\Delta a_{10}^*$ be the actual semi-major after performing its in plane maneuver. Note that $\Delta a_{10}^*$ already includes the possible errors in the maneuver of the master satellite. From the value $\Delta a_{10}$ it is possible to obtain the reference density ($\rho_r$) that was used to define the last maneuver of the master satellite. In particular, for the case in which the cross track maintenance is more restrictive than the along track maintenance:
\begin{equation}
T_M = 4\sqrt{\frac{\Delta\lambda_{s1}bc_1}{3\omega_{\oplus}\rho_{r} }\sqrt{\displaystyle\frac{a_0}{\mu}}} = \frac{2\Delta a_{10}bc_1}{\rho_r\sqrt{\mu a_0}},
\end{equation}
and thus:
\begin{equation}
\rho_r = \displaystyle\frac{3}{4}\frac{\omega_{\oplus}bc_1\Delta a_{10}^2}{\Delta \lambda_{s1}a_0\sqrt{\mu a_0}}.
\end{equation}
That way, the expected time between maneuvers that it has to be imposed to the slave satellite is:
\begin{equation}
T_M = \frac{2\Delta a_{10}^*bc_1}{\rho_r\sqrt{\mu a_0}} = \displaystyle\frac{8}{3}\frac{\Delta\lambda_{s1}a_0\Delta a_{10}^*}{\omega_{\oplus}\Delta a_{10}^2}.
\end{equation}
which already includes the effect of the error in the master's maneuver. It is important to note that this time between maneuvers will not happen in reality as the conditions of the maneuvering cycle will be different from the ones predicted. However, this value allows to define the maneuver for the slave satellite.

Now, we have to define the maneuver of the slave satellite. First, we assume that due to previous errors in the prediction of the cycle conditions, the relative position of the slave satellite differs from the nominal. Let $t_e$ be error in the along track distance that the slave satellite is experiencing at the moment in which the maneuver should be performed. This also represents the amount of along track distance that the satellite has to correct in the next maneuvering cycle. Then, using Equation~\eqref{deltatnominal}:
\begin{equation}
0 = t_e - \displaystyle\frac{3}{2}\frac{\Delta a_{20}}{a_0}T_M + \frac{3}{4}\frac{\rho_r}{bc_1bc_r}\sqrt{\frac{\mu}{a_0}}T_M^2,
\end{equation}
the initial value of the semi-major axis of the slave satellite with respect to its nominal orbit can be obtained:
\begin{equation}
\Delta a_{20} = \displaystyle\frac{1}{2}\frac{\rho_r}{bc_1bc_r}\sqrt{\frac{\mu}{a}}T_M + \frac{2}{3}\frac{a_0t_e}{T_M},
\end{equation}
which can be expressed in terms of the maneuver of the master satellite:
\begin{equation}
\Delta a_{20} = \displaystyle\frac{\Delta a_{10}^*}{bc_r} + \frac{2}{3}\frac{a_0t_e}{T_M},
\end{equation}
and for the case of being the cross track the most restrictive boundary:
\begin{equation}
\Delta a_{20} = \displaystyle\frac{\Delta a_{10}^*}{bc_r} + \displaystyle\frac{1}{4}\frac{t_e\omega_{\oplus}\Delta a_{10}^2}{\Delta \lambda_{s1}\Delta a_{10}^*}.
\end{equation}

One additional thing to notice is that the maneuver of the slave satellite will also have errors in its execution. However, this errors will be corrected in the next planned maneuver, as they will be part of the term $t_e$ in the new cycle.

%%%%%%%%%%%%%%%%%%%%%%%%%%%%%%%%%%%%%%%%%%%%%%%%%%%%%%%%%%%%%%%%%%%%%%%%%%%%%%%%%%%%%%%%%%%%%%
%%%%%%%%%%%%%%%%%%%%%%%%%%%%%%%%%%%%%%%%%%%%%%%%%%%%%%%%%%%%%%%%%%%%%%%%%%%%%%%%%%%%%%%%%%%%%%

\subsection{Alternative maneuvering strategy}

In many space applications, specially in Earth observation missions, it is extremely interesting to bound the along track time distance between satellites as much as possible, since that situation allows satellites to provide a more homogeneous quality in their measurements. For that reason, in this section we introduce an alternative master-slave maneuvering strategy in order to reduce the maximum variation in the along track distance between both satellites during their dynamic. This alternative strategy is based on the compromise that even if one of the satellites of the formation is unable to perform its orbital maneuvers, the minimum distance between both satellites remains controlled and bounded at any moment in the dynamic. 

Since the majority of slave satellites usually are smaller than their masters, we will assume that the slave satellite has a smaller ballistic coefficient. Moreover, and without loss in generality, we will assume that the slave satellite flies ahead of his master. Note that other configurations can also be studied using the general formulation from previous sections.

\begin{figure}[h!]
	\centering
	\includegraphics[width=0.6\textwidth]{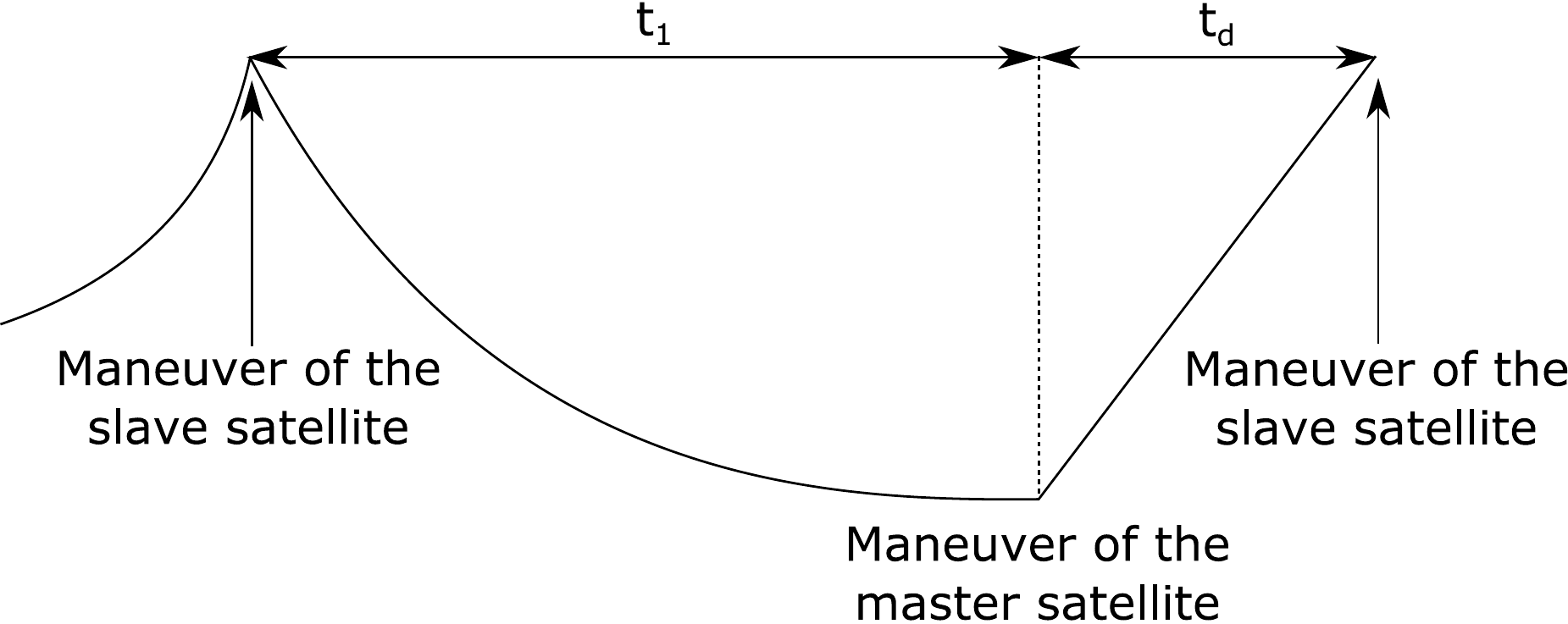} 
	\caption{Multiple maneuvering strategy concept.}
	\label{fig:multiple2}
\end{figure}

In order to fulfill the goals considered, we impose that, whenever possible, both satellites have to present the same semi-major axis when the master satellite performs its maneuver. In addition, the slave satellite will perform its maneuvers when the along track distance matches the maximum of the maneuvering cycle. That way, if the master satellite fails its in plane maneuver, both satellites will separate each other naturally, thus, improving the safety of the mission for such cases. A representation of this dynamic between the last two consecutive maneuvers of the slave satellite can be seen in Figure~\ref{fig:multiple2}, where $t_1$ has been defined as the time from the last additional maneuver of the slave satellite to the maneuver of its master.

From the derivative of Equation~\eqref{dt}, the time defined between the additional maneuver of the slave satellite and the master maneuver ($t_1$) can be computed:
\begin{equation} \label{mul1}
t_1 = \displaystyle\frac{\Delta a_{21} - \Delta a_{11}}{\frac{\rho}{bc_1}\sqrt{\mu a}\left(\frac{1}{bc_r} - 1\right)},
\end{equation}
where:
\begin{equation}\label{mul2}
\Delta a_{11} = \displaystyle\frac{\rho}{bc_1}\sqrt{\mu a}\left(t_1 - \frac{T_M}{2}\right).
\end{equation}
On the other hand, since the initial and final along track distances when the slave satellite performs its maneuvers is the same, the following expression can be derived:
\begin{eqnarray}\label{mul3}
0 & = & - \displaystyle\frac{3}{2}\frac{\Delta a_{21}t_1 + \Delta a_{21}t_d - \frac{\rho}{bc_1bc_r}\sqrt{\mu a}t_1t_d - \Delta a_{11}t_1 - \frac{\rho}{bc_1}\sqrt{\mu a}\frac{T_M}{2}}{a} + \nonumber \\
& + & \frac{3}{4}\frac{\rho}{bc_1}\sqrt{\frac{\mu}{a}}\left(\frac{1}{bc_r} - 1\right)\left(t_1^2 + t_d^2\right).
\end{eqnarray}
Finally, from the system given by Equations~\eqref{mul1},~\eqref{mul2} and~\eqref{mul3}, parameters $t_1$ and $\Delta a_{20}$ can be obtained:
\begin{eqnarray}
t_1 & = & \sqrt{t_d\left(t_d + \displaystyle\frac{2T_M}{\left(\frac{1}{bc_r} - 1\right)}\right)}, \nonumber \\
\Delta a_{21} & = & \displaystyle\frac{\rho}{bc_1}\sqrt{\mu a}\left[\frac{1}{bc_r}t_1 - \frac{T_M}{2}\right].
\end{eqnarray}
It is important to note that if $t_1 > T_M - t_d$ or $bc_r = 1$, it is not possible to include an additional maneuver of the slave satellite under these conditions. Moreover, there are situations where more than one additional maneuver can be included. In such cases, and in order to limit the maximum variation of the along track distance between satellites, the minimum along track distance between the satellites in the period in between the slave maneuvers must be equal to the along track distance between satellites when the master satellite performs its maneuver. That is, by using Equation~\eqref{max}:
\begin{equation}
\Delta\left(\Delta t_{max}\right) = - \displaystyle\frac{3}{4}\frac{\rho}{bc_1}\sqrt{\frac{\mu}{a}}\left(\frac{1}{bc_r} - 1\right)t_1.
\end{equation}
That way:
\begin{eqnarray}
t_i & = & 2t_1, \nonumber \\
\Delta a_{2i} & = & \Delta a_{1i} + \displaystyle\frac{\rho}{bc_1}\sqrt{\mu a}\left(\frac{1}{bc_r} - 1\right)t_1
\end{eqnarray} 
where $t_i$ represents the time between two consecutive maneuvers of the slave satellite, $\Delta a_{2i}$ the initial semi-major axis of the slave satellite with respect to its nominal orbit after its maneuver is performed, and $\Delta a_{1i}$ is the semi-major axis of the master satellite with respect to its reference orbit for the same instant.

%%%%%%%%%%%%%%%%%%%%%%%%%%%%%%%%%%%%%%%%%%%%%%%%%%%%%%%%%%%%%%%%%%%%%%%%%%%%%%%%%%%%%%%%%%%%%%
%%%%%%%%%%%%%%%%%%%%%%%%%%%%%%%%%%%%%%%%%%%%%%%%%%%%%%%%%%%%%%%%%%%%%%%%%%%%%%%%%%%%%%%%%%%%%%
%%%%%%%%%%%%%%%%%%%%%%%%%%%%%%%%%%%%%%%%%%%%%%%%%%%%%%%%%%%%%%%%%%%%%%%%%%%%%%%%%%%%%%%%%%%%%%

\section{Example of application}

In this section we apply the formulation presented in this work to a particular problem. To that end, we first assess the control strategy of a single satellite, and later we study a tandem formation where the previous satellite will be used as the master of the formation. In this example, and without loss in generality, we use sun-synchronous orbits since these are very common in the LEO region. 

First of all, we require to define the mass properties of the satellite in study and its nominal orbit. To that end, we select a satellite of $1285$ $kg$ of mass, cross section of $8.5$ $m^2$, and a drag coefficient of $2.2$, that will fly in a repeating ground-track orbit that repeats its cycle each 27 days or 385 orbital revolutions ($a = 7177.926$ km), with sun-synchronous inclination ($i = 98.602^{\circ}$) and frozen eccentricity ($e = 0.001148$ and $\omega = 90^{\circ}$), with local time at the ascending node of 22:00, and a dead band of $\pm 1$ $km$. This corresponds to a satellite similar to Sentinel-3A, an Earth observation satellite from the European Space Agency.  

\begin{figure}[h!]
	\centering
	\includegraphics[width=0.8\textwidth]{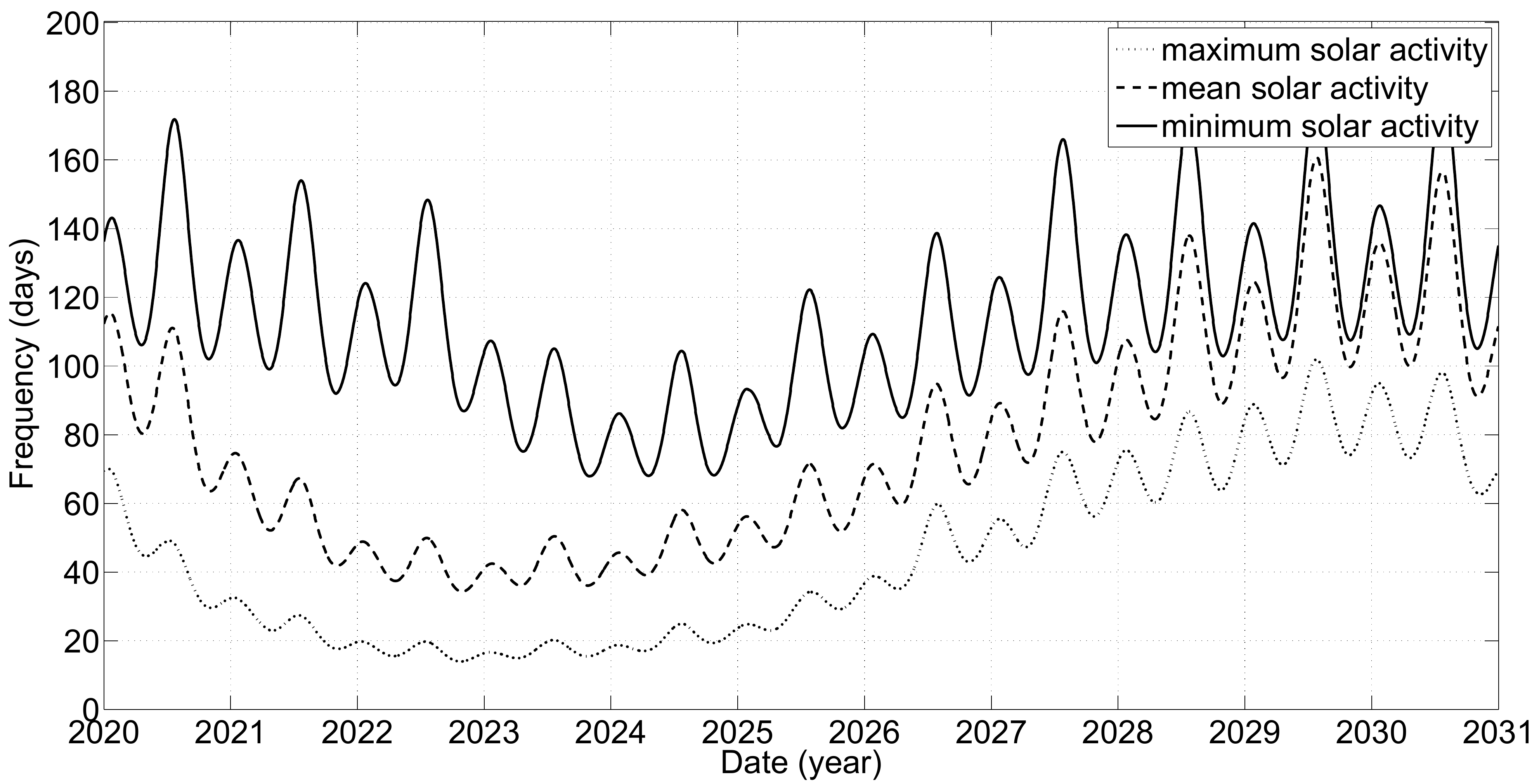} 
	\caption{Maneuver frequency.}
	\label{fig:frequency}
\end{figure}

\begin{figure}[h!]
	\centering
	\includegraphics[width=0.8\textwidth]{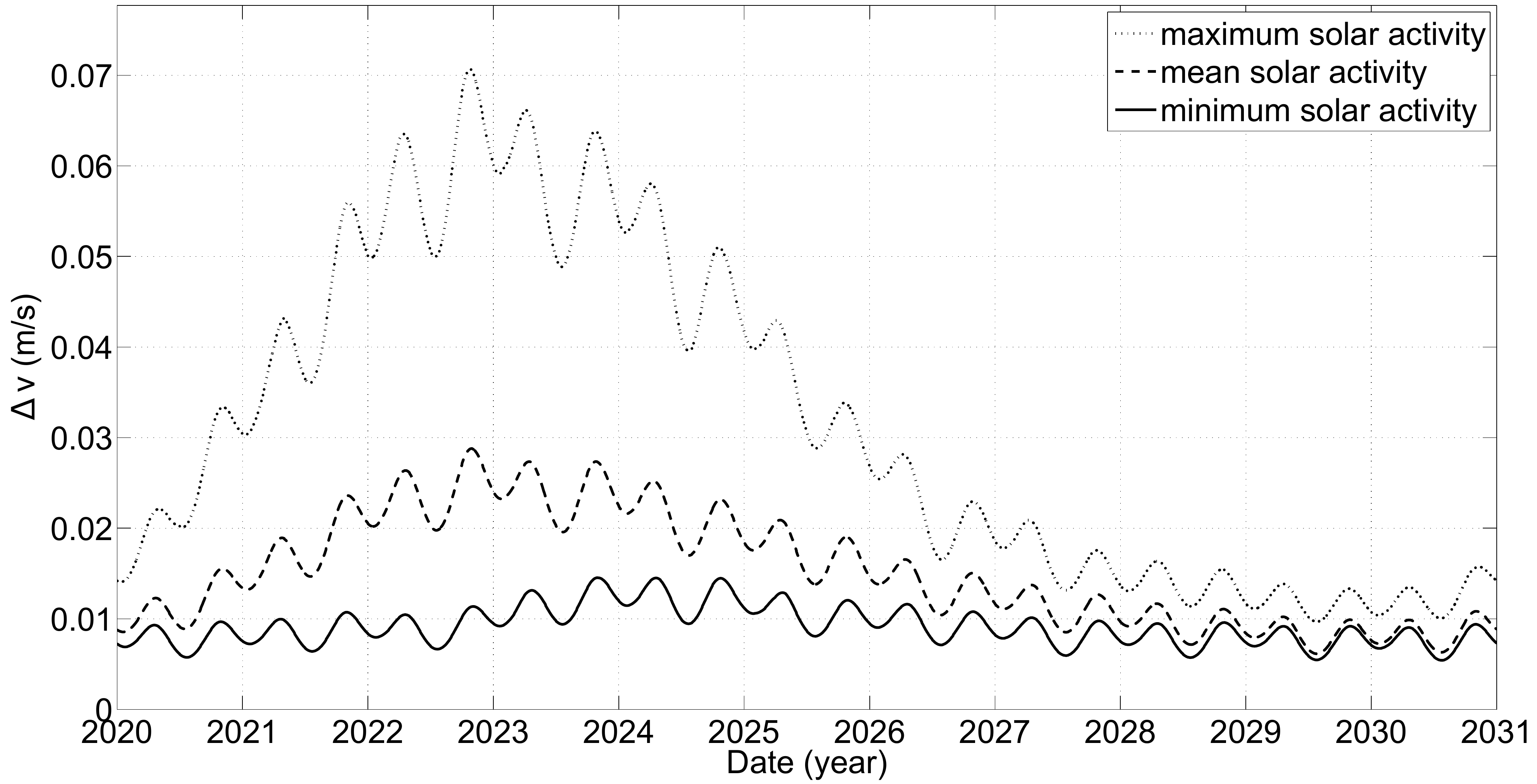} 
	\caption{Impulse per maneuver.}
	\label{fig:impulse}
\end{figure}

Figures~\ref{fig:frequency} and~\ref{fig:impulse} show the frequency and the impulse per maneuver of the satellite for a complete 11 solar cycle including the cases for maximum, mean, and minimum solar activity expected. In these figures, the value of the atmospheric density required for the model was computed using the NRLMSISE-00 model~\cite{msis} for the atmospheric density, and tabulated data from the European Cooperation for Space Standardization (ECSS)~\cite{ecss} for the flux generated in a solar cycle. Comparing this methodology against Vallado's algorithm for satellite station keeping~\cite{Vallado} we have observed relative errors in the order of $0.1\%$ when computing these quantities and for the cases studied. Note that these results are coherent with the simplifications performed in this first order approximation model.

\begin{figure}[h!]
	\centering
	\includegraphics[width=0.8\textwidth]{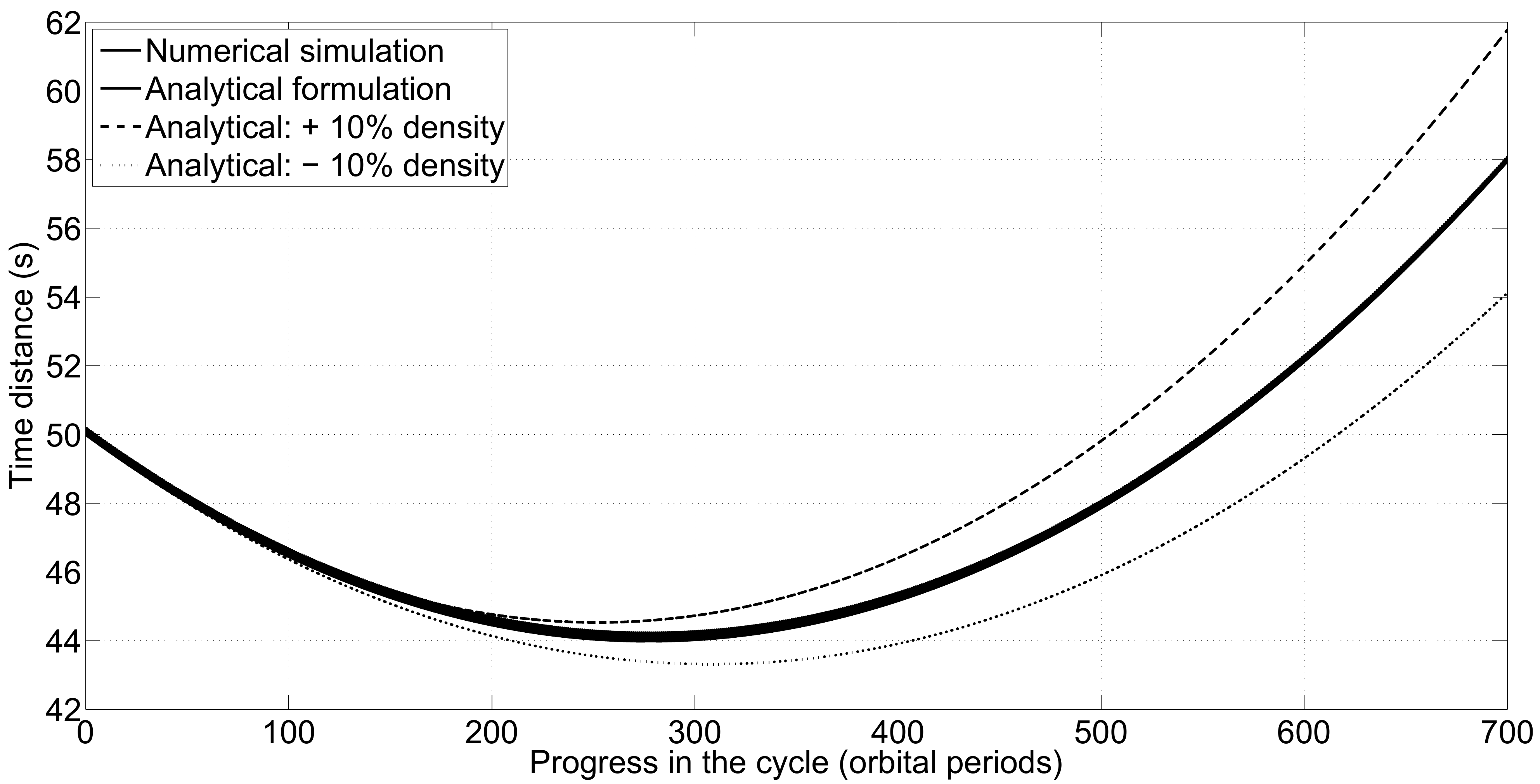} 
	\caption{Along track time distance between satellites.}
	\label{fig:difference}
\end{figure}

Once the master satellite is defined, we can now evaluate the dynamic of a master-slave formation. To that end, we select a satellite that will fly ahead of its master with a different ballistic coefficient ($bc_r = 0.5$) and that will perform its maneuvers three days after the master satellite of the formation. Figure~\ref{fig:difference} shows the evolution of the along track distance between both satellites for a mean density of $\rho = 2.624e-14 kg/m^3$ using both the analytical model presented in this manuscript and a numerical propagator based on a Runge-Kutta method (solid lines) where the perturbations considered are the atmospheric drag and a 4x4 Earth gravitational potential. As can be seen, the analytical and the numerical solutions for the mean density are nearly matching.

Another interesting study is to evaluate how a change in the atmospheric density affects the dynamic of the system. In that sense, Figure~\ref{fig:difference} also shows the along track time distance evolution of the two satellites if a variation of $\pm 10\%$ of the density is considered. As can be observed, the variation of the dynamic due to a change of density is much larger than the effect of the simplifications performed to derive the analytical model. This means that the proposed model can be used not only to define a boundary for the solutions by imposing a level of uncertainty for the density in the analytical model, but also to have a first order solution that is very easy to compute on board spacecrafts.

Additionally, similar experiments have been performed with non constant atmospheric densities. In those cases, instead of using the integral formulation (which assumes a constant mean density), it is possible to integrate the differential formulation (see Equations~\eqref{eq:deltatgeneral} and~\eqref{eq:diftandem}) where now, the atmospheric density affecting the satellite is a function of time and position. As before, we compared this methodology with a direct propagation of the satellites involved obtaining similar results as the ones obtained in Figure~\ref{fig:difference}. Therefore, we show that this methodology can be used as a tool to obtain a first order approximation of the relative motion of the system in a very simple manner.

The model presented in this work has also been successfully applied to the European Space Agency's mission FLEX~\cite{carbon}, which is planned to have a tandem formation with a satellite already in orbit, Sentinel-3. A complete parametric study was performed for that mission using this model~\cite{Flex}, where all results were cross checked with standard orbital propagators. Note also that this model can be used to study the dynamic during the station keeping in plane maneuvers as shown in Ref~\cite{Flex}. Other examples of application of this model for station keeping can be seen in Arnas~\cite{tesis}.

\section{Conclusions}

This work introduces a linearized analytical model to study the dynamic of satellites in near circular Low Earth Orbits under the effects of the atmospheric drag. The model is based on the idea of defining a nominal orbit that already contains the effects of the Earth gravitational potential, and determine the relative motion of the satellites with respect to this nominal orbit. That way, it is possible to decouple the effects of both the Earth gravitational potential and the atmospheric drag to ease the study of these systems. 

The model proposed in this manuscript is first used to evaluate the station keeping maneuvers that satellites require to maintain their orbit in a defined control box. Then, these results are used to extend its application to tandem formations of satellites flying in the same nominal ground-track and study their absolute and relative dynamic. 

This formulation allows to study the problem in a very simple manner, while providing in addition, a clear understanding on the influence of the different variables in the problem. In that respect, this manuscript shows how the along track dynamic between two satellites, flying in tandem under a master slave scenario, depends primary on the density during the dynamic, the relation between the ballistic coefficients of both satellites and the time delay between the in plane maneuvers of the satellites.

On the other hand, this methodology has been used to define two maneuvering strategies and asses their performance regarding the maximum variation of the along track time distance between the satellites of the formation. In particular, the first strategy presented in based on the idea of making the slave satellite to mimic the in plane maneuvers of its master. This allows a simple control and operation of the mission while maintaining the tandem configuration.

In addition, an alternative maneuvering strategy is presented, where, instead of performing just one in plane maneuver of the slave satellite per each one of the master satellite, several are performed in such a way that the maximum variation of the along track distance is reduced and, at the same time, this distance is controlled and bounded at any time, even in the case when one of the satellites is unable to perform orbital maneuvers.

Finally, it is important to note that due to the simplicity of this model, and the low amount of computations that it requires to provide an approximated solution to the problem, this model could be potentially used for the autonomous computation of orbital maneuvers on board spacecrafts. That way, satellites already in orbit could have a first order approximation of their dynamic and possible orbital maneuvers in a simple and fast process.

%%%%%%%%%%%%%%%%%%%%%%%%%%%%%%%%%%%%%%%%%%%%%%%%%%%%%%%%%%%%%%%%%%%%%%%%%%%%%%%%%%%%%%%%%%%%%%
%%%%%%%%%%%%%%%%%%%%%%%%%%%%%%%%%%%%%%%%%%%%%%%%%%%%%%%%%%%%%%%%%%%%%%%%%%%%%%%%%%%%%%%%%%%%%%
%%%%%%%%%%%%%%%%%%%%%%%%%%%%%%%%%%%%%%%%%%%%%%%%%%%%%%%%%%%%%%%%%%%%%%%%%%%%%%%%%%%%%%%%%%%%%%

\end{document}